# Features of photothermal transformation in porous silicon based multilayered structures

K. Dubyk[1*], L. Chepela[1], P. Lishchuk[1], A. Belarouci[2], D. Lacroix[3], M. Isaiev[3,1]

1 Taras Shevchenko National University of Kyiv, 64/13, Volodymyrska Street, 01601 Kyiv, Ukraine
2 Université de Lyon; Institut des Nanotechnologies de Lyon INL-UMR5270, CNRS, Ecole Centrale de Lyon-Université de Lyon, 69134 Ecully, France
3 Université de Lorraine, CNRS, LEMTA, Nancy, F-54000, France

**Abstract**

This paper is devoted to the study of photothermal transformations in multilayered structures. As a modelled sample, porous silicon with a periodic distribution of the porosity was chosen. The spatial distribution of the optical properties inside the structure was evaluated under Brugmann's approximation. The heat sources arising as a result of electromagnetic radiation absorption in the structure were estimated by solving Maxwell's equations. This allowed us to calculate temperature profiles inside photo-excited sample. For experimental measurements, photoacoustic set-up with a gas-microphone transduction system was chosen to investigate thermal properties of the structure. The results of the photoacoustic response simulation based on the gas-piston model demonstrated an excellent agreement with experiments. This allows a reliable evaluation of the thermal conductivity by fitting the experimental amplitude-frequency photoacoustic signal with the simulated one.

---

[*] Corresponding author. Tel.: +380631833456; E-mail: kateryna.dubyk@gmail.com (Kateryna Dubyk)

Photothermal methods, like Raman spectroscopy and thermoreflectance, are widely used for the study of thermal properties of various nanostructured materials [1–4]. These approaches are based on the investigation of photoinduced thermal perturbation inside the medium. The main advantage of such techniques is the possibility to provide flexible contactless measurements. This is very important for the study of nanostructured systems with ultrathin features. However, there are also drawbacks while using such methods like multiple light reflections. For instance, interference pattern can arise in a thin film supported by a substrate that can significantly alter measurements interpretations. Therefore, it is necessary to perform additional simulations of light propagation in the sample to take into account the spatial distribution of the excitation field inside the structure. Miss-evaluating the latter can lead to additional source of errors which impacts directly the thermal properties' assessment. Improving measurement accuracy can be achieved by using several excitation wavelengths to determine thermal parameters [5,6]. However, in this case, light propagation in the investigated structures should be considered even more carefully. Indeed, it maybe nontrivial when taking into account multiple reflections at the interfaces between different layers[7].

In this paper, we investigated photothermal transformation in multilayered porous silicon-based structures. Specifically, structures with periodically modulated in-depth distribution of porosities have been considered. The well-established gas-microphone photoacoustic (PA) technique was adopted for experimental characterization. Hereafter, we proposed to consider the use of several light sources with different wavelengths for thermal perturbation excitations. This approach allows us to increase the sensitivity of the method when evaluating the thermal conductivity.

The samples have been prepared by electrochemical etching of a 500 µm boron doped p+-type silicon substrate, with a high doping of Boron ($10^{19}$ cm$^{-3}$) and a [100] crystal orientation. The electrolyte used for the etching consisted of a 1:2 mixture of absolute ethanol and hydrofluoric (HF) acid (49 %). The production of multilayers was conducted by controlling the anodization current density while other etching parameters were kept constant. The current density was modulated discretely between 7.5 and 80 mA·cm$^{-2}$, so that a multilayer device made of 30 bilayers was achieved.

Thickness and porosity were controlled by tuning etching time and current density with respect of previously established calibration curves[8]. Etching times of 10s and 4s, corresponding to low and high current densities respectively, were applied (see S1 in Supplementary Materials)[9]. The typical cross-sectional scanning electron microscopy (SEM) image of the studied sample and its schematic view are shown in Fig. 1.

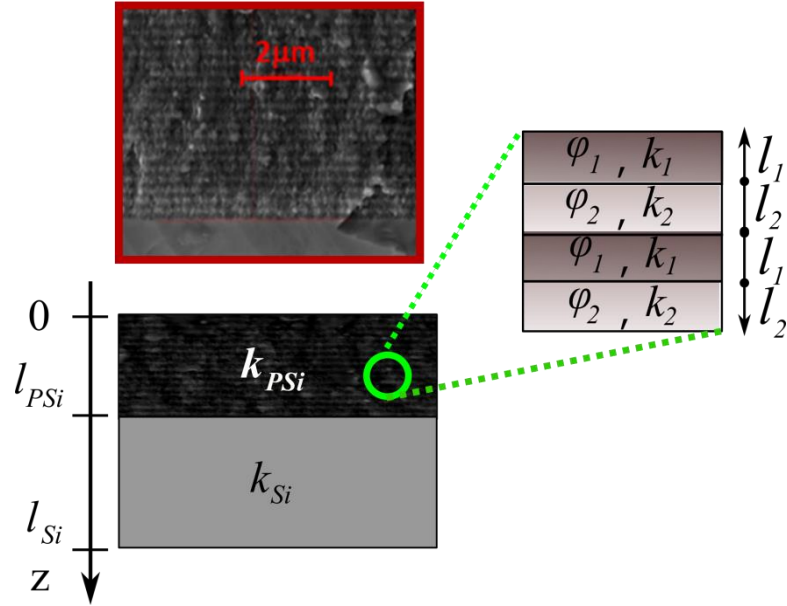

Fig. 1. Cross-sectional SEM image of porous silicon multilayered sample №1 and scheme of multilayered structure.

Thicknesses of porous layers were measured from SEM images. The structural parameters of fabricated samples are presented in Table 1.

Table 1. Structural parameters of the fabricated porous silicon multilayered samples ($\varphi$ is the porosity; $l$ is the thickness; indexes 1, 2 correspond to the first and second porous layer of bilayers, respectively; $l_{PSi}$ and $\rho_{eff}$ are the thickness and the effective density of the multilayered structure)

| № | $\varphi_1$, % | $\varphi_2$, % | $l_1$, nm | $l_2$, nm | $l_{PSi}$, µm | $\rho_{eff}$, kg/m$^3$ |
|---|---|---|---|---|---|---|
| 1 | 46 | 61 | 101 | 115 | 6.48 | 1071 |
| 2 | 44 | 64 | 82 | 136 | 6.54 | 1013 |
| 3 | 46 | 72 | 101 | 192 | 7.98 | 882 |

The following equation was used to calculate the effective density of the multilayered structure:

$$\rho_{eff} = \rho_{SI} \frac{(1-\varphi_1)l_1 + (1-\varphi_2)l_2}{l_1 + l_2} \qquad (1)$$

where $\rho_{SI}$ is the density of bulk silicon.

Photoacoustic study of porous multilayered structures was performed in conventional gas-microphone configuration (see inset in Fig. 2). In such configuration, the sample is in the isolated photo-acoustic chamber, and the top sample surface is illuminated with modulated excitation. Light absorption leads to the heating of the top sample surface and, as a result, of the above gas-layer. The pressure fluctuations induced by this heating were registered with the microphone [10–13].

In this work, laser sources with an output optical power of 100 mW and with different spectral wavelengths of λ = 430 nm; 532 nm; 680 nm were used for the sample illumination. The square-wave modulated irradiation was uniformly distributed over the 25 mm² spot on the 100 mm² sample surface. The spot size and the thermal perturbation penetration length were much bigger than the sample thickness for the considered frequency range. It allows us to use the 1D assumption for thermal transport propagation description.

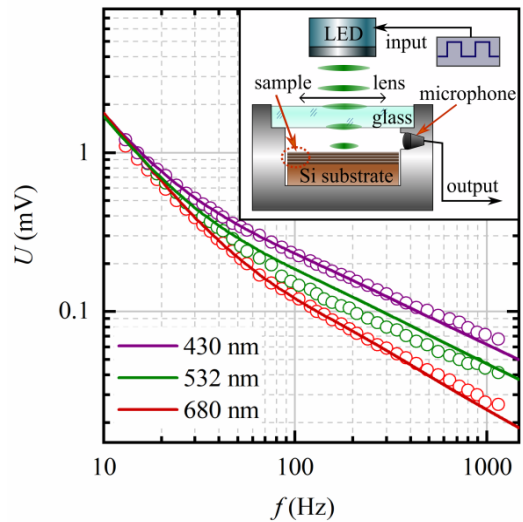

Fig. 2. Experimental amplitude-frequency characteristics of the photoacoustic signal for sample №1 measured in the conventional gas-microphone configuration under laser excitation at different spectral wavelengths: violet - 430 nm; green - 532 nm; red - 680 nm. Circles stands for measurements, solid lines represent the simulation. Inset: scheme of the gas-microphone cell for photoacoustic measurements.

The photoacoustic signal was recorded by an "electret" microphone and compared with the reference provided by the current generator. The amplitude–frequencies dependencies of the photoacoustic signal was measured in the frequency range 20 Hz to 1100 Hz[14]. The amplitude-frequency characteristics of the PA response recorded for samples at different excitation wavelengths are shown in Fig. 2 and in S2 of Supplementary Materials.

For the simulation of the photothermal response, we used Bruggeman's approximation to calculate the complex dielectric permittivity of each porous layer [15,16] (see S3):

$$\varphi \frac{1-\varepsilon_{ps}}{1+2\varepsilon_{ps}} + (1-\varphi)\frac{\varepsilon_b - \varepsilon_{ps}}{\varepsilon_b + 2\varepsilon_{ps}} = 0, \tag{2}$$

where $\varepsilon_b$ and $\varepsilon_{ps}$ are the complex dielectric permittivity of crystalline silicon and porous silicon respectively[8,17].

Furthermore, the spatial reconstruction of the dielectric permittivity ($\varepsilon$) allows us to calculate the in-depth distribution of thermal sources based on Maxwell's equations[18,19]:

$$\vec{\nabla} \times \vec{H} = \sigma \vec{E} + i\omega\varepsilon\vec{E}, \tag{3}$$

$$\vec{\nabla} \times \vec{E} = -i\omega\mu\vec{H}, \tag{4}$$

where $\vec{H}$ is the magnetic field, $\vec{E}$ is the electrical field, $\sigma$ is the electrical conductivity. The volumetric power $P_V$ dissipated in the media (photoinduced heat source) was expressed as:

$$P_V = -Im(\varepsilon)\omega E^2 \tag{5}$$

The spatial distribution of heat sources induced by electromagnetic radiation at different wavelengths (430 nm and 532 nm) are presented in Fig. 3. They were calculated using real and imaginary parts of the refractive index of porous multilayered sample №1 with COMSOL Multiphysics software.

We considered the heating of the sample with the use of a periodic heat source to have a direct comparison with photoacoustic experiments. In this case, the energy balance for the different components of temperature was written as follow:

$$\frac{\partial}{\partial z}\left(k\frac{\partial \theta}{\partial z}\right) - c\rho i\omega\theta = P_V \tag{6}$$

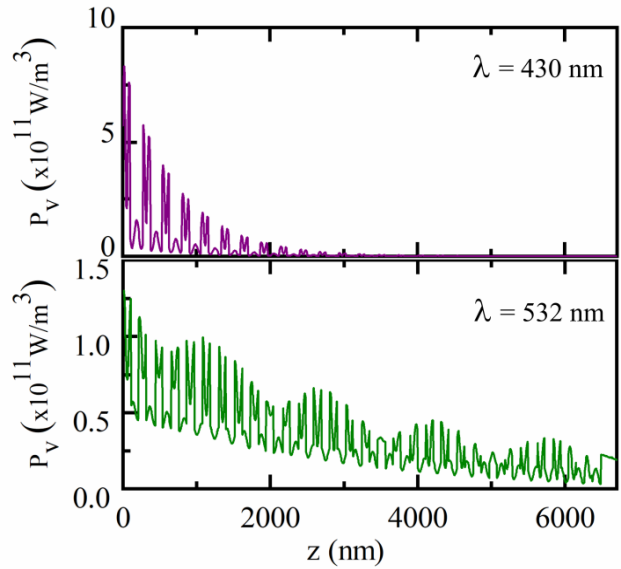

Fig. 3. Spatial distribution of heat sources induced by the electromagnetic radiation at 430 nm and 532 nm.

here $\theta$ is the temperature, $\kappa$ is the thermal conductivity, $c$ is the heat capacity, $\rho$ is the density, $\omega$ is the angular frequency. The spatial distribution of thermal conductivity was considered as follows:

$$k = \begin{cases} k_{eff}, & 0 < z < l_{PSi} \\ k_{Si}, & l_{PSi} < z < l_{Si} \end{cases} \quad (7)$$

where $k_{eff}$ is the effective thermal conductivity of the studied multilayered system and $k_{Si} = 110$ W/(m·K) is the thermal conductivity of highly doped crystalline silicon substrate[20]. $l_{Si}$ is the overall thickness of the system. The Eq. (6) was supplemented with the boundary conditions based on the absence of heat outflows from the top and bottom sample surfaces. The sample was heated as a result of volumetric source occurrence (right-hand-side of Eq. (6)), which was calculated with Eq. (5). We assume continuity conditions of the temperature and heat fluxes through the interface between the multilayered porous structure and the bulk silicon substrate. In addition, it is worth mentioning that the elevation of temperature within the structure is weak, below 1 K. The voltage $U$ recorded by the microphone (gas-microphone detection in classical configuration) was proportional to the pressure $p$ of the "gas piston" in the photoacoustic cell according to the Rosencsweig-Gersho model and was estimated with the following expressions [10–12].

$$U_{sim}(\omega) \sim p \sim \int_0^{-\infty} \theta(0) \exp(\mu_g z)\, dz = \theta(0)/\mu_g, \quad (8)$$

$$\mu_g = \sqrt{i\omega c_g \rho_g / \kappa_g}, \quad (9)$$

here $\theta(0)$ is the temperature at the surface of the sample, $c_g, \rho_g$ and $\kappa_g$ are respectively the heat capacity, density and thermal conductivity of the gas in the photoacoustic cell. The equation (8) was used for the simulation of the amplitude-frequency photoacoustic response.

The simulated amplitude-frequency dependencies for different value of the thermal conductivity are plotted in Fig. 4. The optical parameters were those considered for sample №1 in the case of 532 nm light excitation wavelength.

The merits of a multi-wavelengths approach for thermal conductivity measurements can be understood considering the variations of the parameter G as a function of frequency (see Fig. 5). The parameter G was defined as the amplitude signal normalized by the amplitude of the laser excitation. It reads:

$$G(k_{eff}, \lambda_i, f) = \frac{U(k_{eff}, \lambda_i, f)}{U(k_{eff}, \lambda_g, f)}, \quad (10)$$

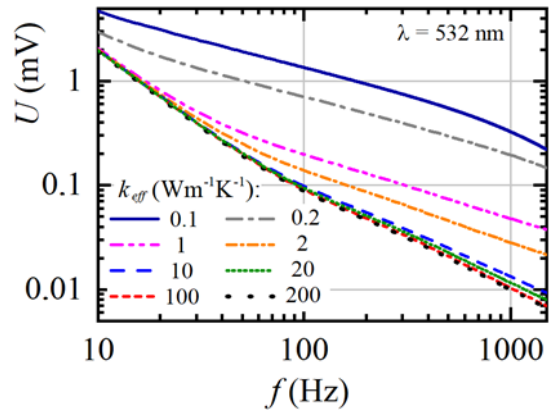

Fig. 4. Amplitude-frequency dependencies for different values of the thermal conductivity.

where $U(k_{eff}, \lambda_i, f)$ is the amplitude frequency dependence of the photoacoustic response excited with light at a wavelength of $\lambda_i$, $k_{eff}$ is the thermal conductivity of the multilayered porous silicon based structure. As one can see, the use of several wavelengths give additional degree of freedom to improve the accuracy for thermal conductivity evaluation. The simulated amplitude-frequency dependencies for the excitation sources with wavelengths from 300 nm to 1000 nm are presented in S4.

The simulated amplitude with Eq. 8 for a given excitation wavelength, was considered as a function of the effective thermal conductivity of the studied multilayered structure. The matching of experimental and simulation results were achieved with the least square method:

$$F(k_{eff}) = \sum_{i=1}^{N}\left(U_{sim}(\omega_i, k_{eff}) - U_{exp}(\omega_i)\right)^2, \quad (11)$$

where $\{\omega_i\}$ is the set of frequencies at which amplitude frequency-dependencies were measured. In our case, we achieved minimization for all excitation sources, considering the following minimization:

$$\frac{\delta(F_g + F_r + F_b)}{\delta k_{eff}} = 0, \quad (12)$$

where $F_g, F_r$ and $F_b$ are the least square sums for the green, red and blue laser sources respectively according to Eq. 11.

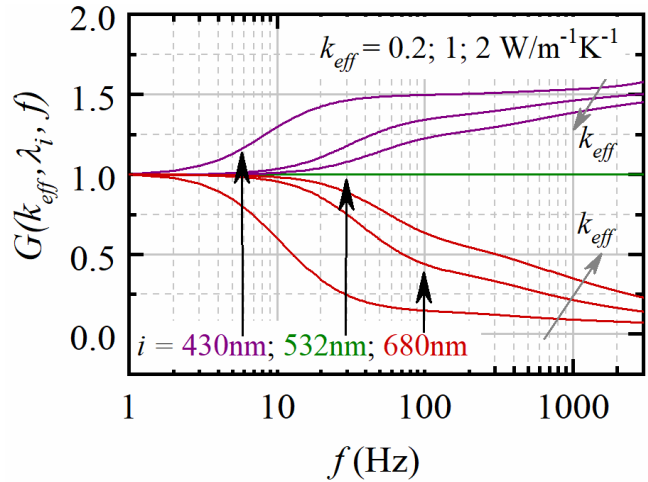

Fig. 5. $G(k_{eff}, \lambda_i, f)$ parameter for different thermal conductivities and for several excitation wavelengths, ($\lambda$ = 430, 532, 680 nm).

With the above discussed technique and the proposed multi-wavelengths model, the effective thermal conductivity of multilayered porous system was extracted. The recovered values are presented in Table 2 for each sample.

The expected thermal conductivity values calculated with the electro-thermal analogy[8] $(l_1 + l_2)/k = l_1/k_1 + l_2/k_2$ (where $l_1$ and $l_2$ are the thicknesses of the first and second porous layer of the bilayer system, while $k_1$ and $k_2$ are the thermal conductivities of such bilayers) based on the well-known thermal conductivity values of porous silicon layers with porosities $\varphi_1, \varphi_2$ respectively[11,21,22] should be between 1.46 and 1.73 W/(m·K) for the investigated samples. However, the experimentally evaluated $k_{eff}$ values (see Table 2) are 2-4 times smaller than the expected ones. The origin of such discrepancies was supposed to be related to the presence of interfacial thermal resistance. The resulting internal resistance was estimated based on the following equation [8]

$$R = \frac{1}{2}\left(\frac{l_1}{k_1} + \frac{l_2}{k_2} - \frac{l_1+l_2}{k_{eff}}\right) \quad (13)$$

Interfacial thermal resistances deduced from Eq. 13 are reported in Table 2.

Table 2. The effective thermal conductivity of porous silicon multilayered structures and the interfacial thermal resistance between porous layers

| № | $k_{eff}$, W/(m·K) | $R$, $10^{-7}$ m²·K/W |
|---|---|---|
| 1 | 0.8±0.1 | 0.73±0.08 |
| 2 | 0.65±0.08 | 1.0±0.1 |
| 3 | 0.40±0.05 | 2.66±0.12 |

In conclusion, we considered the features of photothermal transformation in multilayered porous silicon systems. The experiments were carried out on samples with a periodic distribution of the optical properties depending on the local porosity. The thermal response was investigated with the gas-microphone photoacoustic technique in classical configuration. For the simulation of the photoacoustic response we exploited a bottom-up approach starting with the expression of the dielectric properties of each layer based on the Bruggeman approximation. Then, the volumetric heat source was evaluated in the media based on Maxwell equations resolution. This heat source was used to calculate the temperature distribution inside the media and the characteristics of the photoacoustic response. Finally, the experimental amplitude-frequency dependency was fitted with the developed model to extract the thermal conductivity of the multilayered structures. This approach was successfully applied with several excitation wavelengths and a mean square minimization to improve the accuracy.

**Supplementary Materials**

Supplementary materials contain information about the etching conditions and the experimentally obtained amplitude-frequency characteristics of the photoacoustic signal under three different laser excitation wavelengths. Additionally, it contains the dielectric functions of porous media, the sensibility of the photoacoustic signal characteristics to wavelength of the excitation source simulated by the presented model.

**Acknowledgements**

This work has been partially funded by the CNRS Energy unit (Cellule Energie) through the project ImHESurNaASA. We want to acknowledge the partial financial support of the scientific pole EMPP of University of Lorraine. The publication contains the results obtained in the framework of the research project "Features of photothermal and photoacoustic processes in low-dimensional silicon-based semiconductor systems" (Ministry of Education and Science of Ukraine, state registration number 0118U000242).

**References**

1 Q. Li, K. Katakami, T. Ikuta, M. Kohno, X. Zhang, and K. Takahashi, *Carbon N. Y.* **141**, 92 (2018).

2 N. Peimyoo, J. Shang, W. Yang, Y. Wang, C. Cong, and T. Yu, *Nano Res.* **8**, 1210 (2015).

3 Z. Luo, J. Maassen, Y. Deng, Y. Du, R.P. Garrelts, M.S. Lundstrom, P.D. Ye, and X. Xu, *Nat. Commun.* **6**, 8572 (2015).

4 M. Isaiev, O. Didukh, T. Nychyporuk, V. Timoshenko, and V. Lysenko, *Appl. Phys. Lett.* **110**, 011908 (2017).

5 M. Isaiev, S. Tutashkonko, V. Jean, K. Termentzidis, T. Nychyporuk, D. Andrusenko, O. Marty, R.M. Burbelo, D. Lacroix, and V. Lysenko, *Appl. Phys. Lett*. **105**, 031912 (2014).

6 A.I. Tytarenko, D.A. Andrusenko, A.G. Kuzmich, I. V. Gavril'chenko, V.A. Skryshevskii, M. V Isaiev, and R.M. Burbelo, *Tech. Phys. Lett*. **40**, 188 (2014).

7 I.A. Lujan-Cabrera, C.F. Ramirez-Gutierrez, J.D. Castaño-Yepes, and M.E. Rodriguez-Garcia, *Phys. B Condens. Matter* **560**, 133 (2019).

8 P. Lishchuk, A. Dekret, A. Pastushenko, A. Kuzmich, R. Burbelo, A. Belarouci, V. Lysenko, and M. Isaiev, *Int. J. Therm. Sci*. **134**, 317 (2018).

9 X.G. Zhang, *J. Electrochem. Soc.* **151**, C69 (2004).

10 M. Isaiev, P.J. Newby, B. Canut, A. Tytarenko, P. Lishchuk, D. Andrusenko, S. Gomes, J.M. Bluet, L.G. Frechette, V. Lysenko, and R. Burbelo, *Mater. Lett*. **128**, 71 (2014).

11 P. Lishchuk, D. Andrusenko, M. Isaiev, V. Lysenko, and R. Burbelo, *Int. J. Thermophys*. 36, 2428 (2015).

12 P. Lishchuk, M. Isaiev, L. Osminkina, R. Burbelo, T. Nychyporuk, and V. Timoshenko, *Phys. E Low-Dimensional Syst. Nanostructures* **107**, 131 (2019).

13 A. Assy, S. Gomes, M. Isaiev, and P. Lishchuk, in Nanostructured Semicond. Amorph. Therm. Prop., edited by K. Termentzidis (CRC Press, Boca Raton, 2017), pp. 493–516.

14 M. Isaiev, D. Andrusenko, A. Tytarenko, A. Kuzmich, V. Lysenko, and R. Burbelo, *Int. J. Thermophys.* **35**, 2341 (2014).

15 L. Chen, L. Zhang, Q. Kang, H.S. Viswanathan, J. Yao, and W. Tao, *Sci. Rep*. **5**, 8089 (2015).

16 W.M. Merrill, S. Member, R.E. Diaz, M.M. Lore, M.C. Squires, and N.G. Alexopoulos, *IEEE Transactions on Antennas and Propagation* **47**, 1, 142-148 (1999).

17 D. Stroud, *Superlattices Microstruct*. **23**, 567 (1998).

18 J. Anto and R.C. Thiagarajan, *Proc. 2012 COMSOL Conf. Bangalore*,

19 G. Tiwari, S. Wang, J. Tang, and S.L. Birla, *J. Food Eng*. **105**, 48 (2011).

20 M. Asheghi, K. Kurabayashi, R. Kasnavi, and K.E. Goodson, *J. Appl. Phys*. **91**, 5079 (2002).

21 K. Dubyk, A. Pastushenko, T. Nychyporuk, R. Burbelo, M. Isaiev, and V. Lysenko, *J. Phys. Chem. Solids* **126**, 267 (2019).

22 A. Ould-abbas, M. Bouchaour, and N.C. Sari, *Open Journal of Physical Chemistry*, **2,** 1-6 (2012).

# Supplementary Materials

S1. The etching parameters of the studied samples

| № | $t_1$, s | $t_2$, s | $j_1$, mA·cm$^{-2}$ | $j_2$, mA·cm$^{-2}$ |
|---|---|---|---|---|
| 1 | 10 | 4 | 10 | 40 |
| 2 | 10 | 4 | 7.5 | 50 |
| 3 | 10 | 4 | 10 | 80 |

S2. Experimental amplitude-frequency characteristics of the photoacoustic signal for samples 2 – (a), and 3 – (b) measured in the conventional gas-microphone configuration under the excitation by laser sources with different spectral wavelengths: (violet - 430 nm, green - 532 nm; red - 680 nm circles respectively). Solid lines represent the simulation.

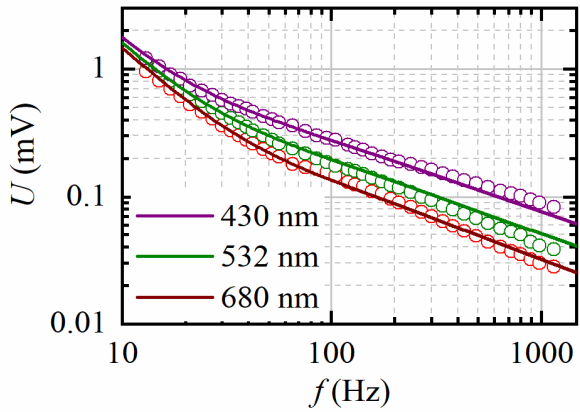
(a)

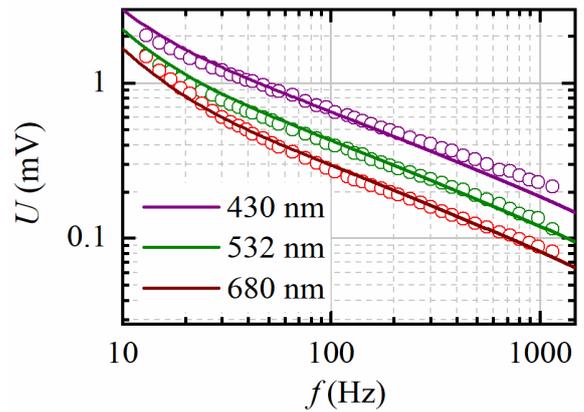
(b)

S3. The real (a) and imaginary (b) part of the complex dielectric permittivity of porous silicon versus the porosity for the different wavelengths used in the experiment.

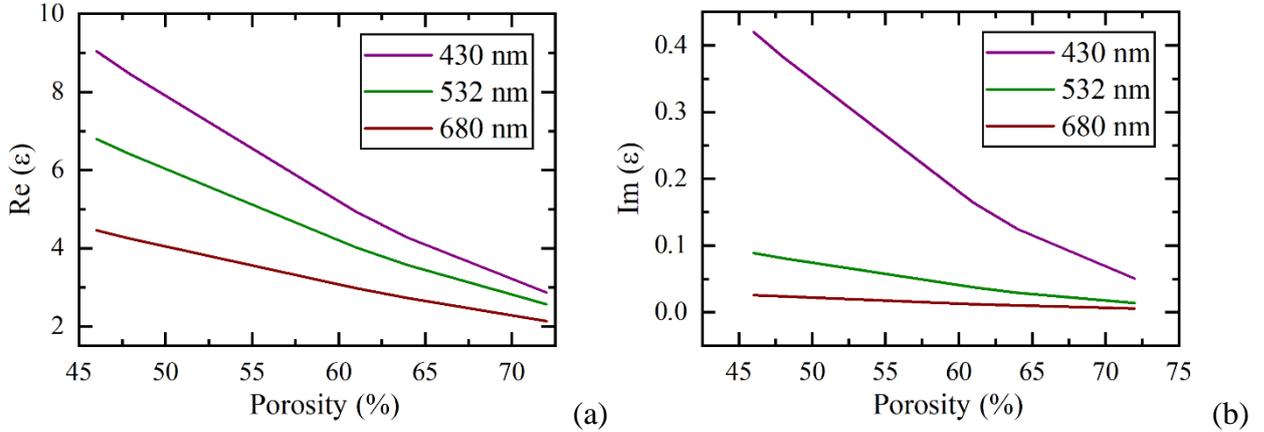

S4. Simulated amplitude-frequency dependencies of the photoacoustic signal for sample 1 under the excitation by laser sources with different spectral wavelengths (400-1000 nm). The dependencies were calculated for the same absorbed energy. We present the curves for the same absorbed energy in order to demonstrate change in general curve shape rather than on the magnitude which often very hard to control in real experimental conditions.

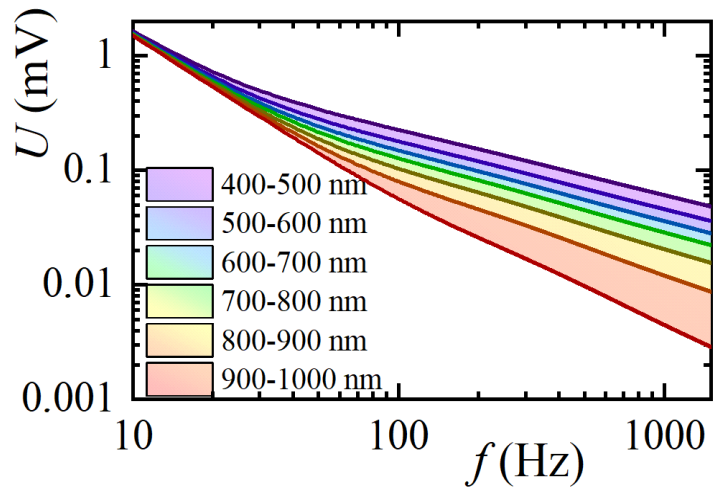